\documentclass[12pt]{article}
\usepackage{amssymb}
\def\F{\mathbb{F}}

\def\Z{\mathbb{Z}}

\usepackage{color,xcolor}
\usepackage{latexsym}
\usepackage{cite}
\usepackage{amsfonts}
\usepackage{amsfonts,mathrsfs}
\usepackage{amsmath,amssymb,amsbsy,amsfonts,amsthm,latexsym,
                        amsopn,amstext,amsxtra,euscript,amscd,color}
\usepackage{cite}
\usepackage{graphicx}
\usepackage{psfrag}
\usepackage{subfigure}
\usepackage{url}
\usepackage{stfloats}
\usepackage{amsmath}

\newtheorem{theorem}{Theorem}
\newtheorem{lemma}{Lemma}

\newcommand{\quash}[1]{}

\begin{document}

\title{Trace representation of pseudorandom binary sequences derived from Euler quotients}

\author{Zhixiong Chen$^{1}$, Xiaoni Du$^{2}$ and Radwa Marzouk$^{3}$\\
1. School of Mathematics, Putian University\\ Putian, Fujian
351100, P. R. China \\
Email: ptczx@126.com\\
2. College of Mathematics and Information Science \\ Northwest
Normal University, Lanzhou, Gansu 730070, P. R. China\\
Email: ymldxn@126.com\\
3. Department of Mathematics, Faculty of Science\\
Cairo University, Giza 12613, Egypt\\
E-mail: radwa@sci.cu.edu.eg}

\maketitle

\begin{abstract}
We give the trace representation of a family of binary sequences derived from Euler quotients by determining
the corresponding defining polynomials. Trace representation can help us producing the sequences efficiently and analyzing
their cryptographic properties, such as linear complexity.
\newline
\textbf{Keywords.} Cryptography; Pseudorandom binary sequences; Euler quotients; Fermat quotients; Trace function.
\newline
{\bf MSC(2010):} 94A55, 94A60, 65C10, 11B68
\end{abstract}

\section{Introduction}

For an odd prime $p$,  integers $r\geq 1$ and  $u$ with
$\gcd(u,p)=1$, the {\it Euler quotient\/} modulo $p^r$, denoted by $Q_{r}(u)$,
is defined as the unique integer by
$$
Q_{r}(u) \equiv \frac{u^{\varphi(p^r)} -1}{p^r} ~(\bmod ~p^r),
\quad 0 \le Q_{r}(u) \le p^r-1,
$$
where $\varphi(-)$ is the Euler totient function. See, e.g., \cite{ADS,CW11,Sha} for details. In addition, we define
$$
Q_{r}(u) = 0  \quad \mathrm{if}~ p|u.
$$

It is easy to verify
\begin{equation}\label{eq:add struct1}
Q_{r}(uv)=Q_{r}(u)+Q_{r}(v) \pmod {p^r},~\gcd(uv,p)=1
\end{equation}
and
\begin{equation}\label{eq:add struct2}
Q_{r}(u+kp^{r})\equiv Q_{r}(u) - kp^{r-1}u^{-1} \pmod {p^r}, ~ \gcd(u,p) = 1, ~ k\in\mathbb{Z}.
\end{equation}
In particular, $Q_{1}(u)$ is called the {\it Fermat quotient}. Many number theoretic have been studied for Fermat and Euler quotients
in \cite{ADS,BFKS,C,CW2,CW3,CW11,ErnMet2,OS,Sha,S,S2010,S2011,S2011b,SW} and references therein.

More recently, Fermat and Euler quotients are studied from the viewpoint of
cryptography, see \cite{C2014,CD,CG,CHD2011,CNW,COW,DCH-IPL,DKC,GW,OS}. Families of pseudorandom sequences are derived from  Fermat and Euler quotients.

In this correspondence, we still concentrate on a family of binary sequences $(e_u)$ defined by Euler quotients. For fixed $\mathfrak{r}\ge 1$,
$(e_u)$ is defined as
\begin{equation}\label{binarythreshold}
e_u=\left\{
\begin{array}{ll}
0, & \mathrm{if}\,\ 0\leq Q_{\mathfrak{r}}(u)/p^{\mathfrak{r}}< \frac{1}{2},\\
1, & \mathrm{if}\,\ \frac{1}{2}\leq Q_{\mathfrak{r}}(u)/p^{\mathfrak{r}}< 1,
\end{array}
\right. ~~~ u\ge 0.
\end{equation}
We note that $(e_u)$ is $p^{\mathfrak{r}+1}$-periodic by (\ref{eq:add struct2}).
The linear complexity of $(e_u)$ is investigated in \cite{CD} for $\mathfrak{r}=1$ and in \cite{DCH-IPL} for $\mathfrak{r}>1$, respectively.
Here, we will investigate a way to produce such binary sequences using trace function, which is extensively applied to producing pseudorandom sequences efficiently and analyzing
their pseudorandom properties \cite{GG}. In particular, in \cite{C2014} the first author has studied the trace representation of $(e_u)$ for $\mathfrak{r}=1$, the idea of which
 helps us to consider the case of $\mathfrak{r} \ge 2$.

We organize this correspondence as follows. In Section 2, we introduce generalized cyclotomic classes of $\mathbb{Z}_{p^{r}}$ by using Euler quotients
and determine the defining pair (see below for the definition) of $(e_u)$. In Section 3, we present the trace representation of $(e_u)$ in terms of its defining pair.
We also give some remarks on the relationship between the defining pair of $(e_u)$ and its linear complexity in the last section.

We conclude this section by introducing the definition of defining pair of a binary sequence.
Let $\mathbb{F}_2=\{0,1\}$ be the binary field and $\overline{\F}_2$ the algebraic closure of $\mathbb{F}_2$.
For  a binary sequence $(s_u)$ over $\F_2$ of odd period $T$, there exists a primitive $T$-th root
$\beta\in\overline{\F}_2$ of unity and a polynomial $G(x) \in \overline{\F}_2[x]$ of degree smaller than $T$ such that
$$
s_u=G(\beta^u), ~~ u\ge 0,
$$
see \cite[Theorem 6.8.2]{J}, we call the pair  $(G(x),\beta)$ a \emph{defining pair} of $(s_u)$ and $G(x)$
the \emph{defining polynomial} of $(s_u)$ corresponding to $\beta$  \cite{DGS,DGS09,DGSY}. Note that for a
 given $\beta$, $G(x)$ is uniquely determined up to modulo $x^{T}-1$ \cite[Lemma 2]{DGSY}.

\section{Defining pair}

We denote by
$\mathbb{Z}_{m}=\{0,1,\ldots, m-1\}$ the residue class ring
modulo $m$ and by $\mathbb{Z}_{m}^*$ the unit group of
$\mathbb{Z}_{m}$. According to (\ref{eq:add struct1}) and (\ref{eq:add struct2}), the quotient $Q_r(-)$ defines a group epimorphism from $\mathbb{Z}^*_{p^{r+1}}$ to $\mathbb{Z}_{p^{r}}$.

Let
$$
D_l^{(r)}=\{u: 0\le u\le p^{r+1}-1,~ \gcd(u,p)=1,~ Q_{r}(u)=l\}
$$
for $l=0,1,\ldots,p^r-1$. Clearly, $D_0^{(r)}, D_1^{(r)},\ldots,
D_{p^r-1}^{(r)}$ form a partition of $\mathbb{Z}_{p^{r+1}}^*$.

Since $\mathbb{Z}^*_{p^{r+1}}$ is cyclic, we choose the element $g\in\mathbb{Z}^*_{p^{r+1}}$ as a
 generator ($g$ is also called a primitive  element of $\mathbb{Z}^*_{p^{r+1}}$).
 We note here that the order of $g$, i.e., the least positive number $n$ satisfying
 $g^n\equiv 1 \pmod {p^{r+1}}$, is $\varphi(p^{r+1})$.
For convenience, we will choose a primitive element $g$ such that $Q_{r}(g)=1$.
One might ask whether such $g$ exists or not? In fact, we suppose that $Q_{r}(g)=a\neq 1$. It is easy to prove that $\gcd(a,p)=1$. By (\ref{eq:add struct2}) we get
$Q_{r}(g^{a^{-1}})=1$, where $a^{-1}$ is the inverse of $a$ modulo $p^{r}$. Furtherly, we have
$$
Q_r(g^{a^{-1}+kp^r})\equiv 1 \pmod {p^{r}}
$$
for all $0 \le k < p-1$. One can find a $k_0~ (0 \le k_0 < p-1)$ such that $\gcd(a^{-1}+k_0p^r, \varphi(p^{r+1}))=1$, i.e.,
$g^{a^{-1}+k_0p^r}$ is primitive modulo $p^{r+1}$ and $Q_{r}(g^{a^{-1}+k_0p^r})=1$. Then we choose $g^{a^{-1}+k_0p^r}$
instead of $g$.

From now on, we always suppose that $Q_{r}(g)=1$ for a fixed primitive element $g$  modulo $p^{r+1}$. By (\ref{eq:add struct1}) we get
$$
D_0^{(r)}=\{g^{kp^r} \pmod {p^{r+1}} : ~0 \le k < p-1 \}
$$
and
$$
D_l^{(r)}:=g^{l} D_0^{(r)}= \{g^{l+kp^r} \pmod {p^{r+1}}: ~0 \le k < p-1 \}
$$
for $1\le l< p^r$. So each $D_l^{(r)}$ exactly contains $p-1$ many elements. We will use the notation $D_{l+p^r}^{(r)}=D_l^{(r)}$ in the context.

Let $\mathcal{I}=\{(p^\mathfrak{r}+1)/2, (p^\mathfrak{r}+3)/2, \ldots, p^\mathfrak{r}-1\}$,
one can define
$(e_u)$ equivalently by
\begin{equation}\label{other-def}
e_u=\left\{
\begin{array}{ll}
1, & \mathrm{if}\,\ u\bmod {p^{\mathfrak{r}+1}}\in \cup_{l\in\mathcal{I}}D_l^{(\mathfrak{r})},\\
0, & \mathrm{otherwise},
\end{array}
\right. \quad u\ge 0,
\end{equation}
which helps us to determine the defining pair and hence the trace representation.

\begin{lemma}\label{lemma1}
For $r\geq 1$, let $uD_l^{(r)}=\{uv \pmod {p^{r+1}}: v\in
D_l^{(r)}\}$. If $u\in D_{l'}^{(r)}$, then we have
$$
uD_{l}^{(r)} = D_{l+l' ~(\bmod~ p^r)}^{(r)},
$$
where $0\leq l, l' < p^r$.
\end{lemma}
Proof. The desired result follows from (\ref{eq:add struct1}).
~ \hfill $\square$

\begin{lemma}\label{lemma02}
For $r\geq 1$ and $0\leq  l < p^{r+1}$,  we have
$$
\{u \pmod {p^{r+1}} : u\in D_{l}^{(r+1)} \} =D_{l \pmod
{p^r}}^{(r)}.
$$
\end{lemma}
Proof.  For an integer $u$ with $p\nmid u$, we write by Euler's theorem
$$
u^{\varphi(p^r)}=1+\ell p^r\in\mathbb{Z},
$$
where $\ell=\ell_0+\ell_1 p^r+\ell_2 p^{2r}+\ldots \in\mathbb{Z}$ with $0\le \ell_i<p^r$ for $i\ge 0$. Then by the definition of Euler quotients, we
have $Q_r(u)\equiv \ell \equiv \ell_0 \pmod {p^r}$. On the other hand, $Q_{r+1}(u)\equiv \ell+\frac{p-1}{2}\ell^2p^{r} \equiv \ell_0 +(\frac{p-1}{2}\ell_0^2+\ell_1)p^r\pmod {p^{r+1}}$,
which is deduced from
$$
u^{\varphi(p^{r+1})}=(u^{\varphi(p^r)})^p=(1+\ell p^r)^p=1+\ell p^{r+1}+\frac{p-1}{2}\ell^2p^{2r+1}+\ldots.
$$
Therefore, we derive
$$
Q_{r+1}(u)\equiv Q_{r}(u) \pmod {p^r},
$$
which leads to
$$
\{u \pmod {p^{r+1}} : u\in D_{l}^{(r+1)} \} \subseteq D_{l \pmod
{p^r}}^{(r)}.
$$

Now we show the cardinality of $ \{u \pmod {p^{r+1}} : u\in
D_{l}^{(r+1)} \}$ is $p-1$, wich equals that of $D_{l \pmod
{p^r}}^{(r)}$. In fact, if $u \equiv u' \pmod {p^{r+1}}$  for
$u,u'\in D_{l}^{(r+1)}$, we suppose $u'=u+k_0p^{r+1}$ for some
$0\le k_0<p$. We have
\begin{equation*}
\begin{split}
l & \equiv  Q_{r+1}(u)\equiv Q_{r+1}(u') \equiv
Q_{r+1}(u+k_0p^{r+1})\\
&\equiv Q_{r+1}(u)-k_0u^{-1}p^{r} \pmod {p^{r+1}},
\end{split}
\end{equation*}
which indicates that $k_0=0$ and hence $u=u'$. We finish the proof. ~ \hfill $\square$\\

Define
$$
D_l^{(r)}(x)= \sum\limits_{u\in D_l^{(r)}}x^u \in
\mathbb{F}_2[x]
$$
for $l=0,1, \ldots, p^r-1$.

For an element $\gamma\in \overline{\mathbb{F}}_{2}$, we denote by $\mathrm{ord}(\gamma)$ the order of $\gamma$, i.e., the least positive integer $n$ such that
$\gamma^n=1$.

\begin{lemma}\label{lemma3}
Let $\gamma\in \overline{\mathbb{F}}_{2}$ be of order $\mathrm{ord}(\gamma)$ with $\mathrm{ord}(\gamma)|p^{r+1}$.
We have
$$
\sum\limits_{l=0}^{p^{r}-1}D_l^{(r)}(\gamma)=\left\{
\begin{array}{ll}
1, & \mathrm{if}\,\ \mathrm{ord}(\gamma)=p,\\
0, & \mathrm{otherwise}.
\end{array}
\right.
$$
\end{lemma}
Proof. If $\mathrm{ord}(\gamma)=1$, i.e., $\gamma=1$, we have
$$
\sum\limits_{l=0}^{p^{r}-1}D_l^{(r)}(1)=p^r(p-1)=0.
$$
Since
$$
  \sum\limits_{l=0}^{p^{r}-1}D_l^{(r)}(\gamma)= \sum_{i\in\mathbb{Z}^*_{p^{r+1}}}\gamma^i=\sum_{i\in\mathbb{Z}_{p^{r+1}}}\gamma^i-\sum_{i\in\mathbb{Z}_{p^{r}}}\gamma^{ip},
 $$
if $\mathrm{ord}(\gamma)>p$, using the formula $1+x+\ldots+x^{n-1}=(1-x^n)/(1-x)$, we have
$$
\sum_{i\in\mathbb{Z}_{p^{r+1}}}\gamma^i=\frac{1-\gamma^{p^{r+1}}}{1-\gamma}=0, ~~~ \sum_{i\in\mathbb{Z}_{p^{r}}}\gamma^{ip}=\frac{1-\gamma^{p^{r+1}}}{1-\gamma^p}=0.
$$
While if $\mathrm{ord}(\gamma)=p$, we have
$$
 \sum_{i\in\mathbb{Z}_{p^{r}}}\gamma^{ip}= \sum_{i\in\mathbb{Z}_{p^{r}}}1=p^{r}=1.
$$
We finish the proof. ~\hfill $\square$\\

 For $r\ge 1$, we define $p^r$-tuples
$$
\mathcal{C}_i^{(r)}(x)=(D_{i}^{(r)}(x),D_{i+1}^{(r)}(x),\ldots,D_{i+p^r-1}^{(r)}(x)), ~~ i=0,1,\ldots,p^r-1.
$$
We also use the notation $\mathcal{C}_i^{(r)}(x)^{\mathrm{T}}$, the transpose of $\mathcal{C}_i^{(r)}(x)$.
We  will calculate the inner product $\mathcal{C}_i(x)\cdot \mathcal{C}_j(x^{p^{m}})^{\mathrm{T}}$ for $0\le i,j<p^r$.

\begin{lemma}\label{inner-product}
Let $\theta \in \overline{\mathbb{F}}_{2}$ be a  primitive
$p^{r+1}$-th root of unity. For any fixed pair $0\le i,j<p^r$, we have
$$
\mathcal{C}_i^{(r)}(\theta)\cdot \mathcal{C}_j^{(r)}(\theta^{p^{m}})^{\mathrm{T}}=0, ~~ r\ge 1
$$
if $m\ge 1$, and
$$
\mathcal{C}_i^{(r)}(\theta)\cdot \mathcal{C}_j^{(r)}(\theta)^{\mathrm{T}}=\left\{
\begin{array}{ll}
1, & \mathrm{if}\,\ p^{r-1}||(i-j),\\
0, & \mathrm{otherwise},
\end{array}
\right. ~~ r\ge 2,
$$
where $p^{r-1}||(i-j)$ means $p^{r-1}|(i-j)$ but $p^{r}\nmid (i-j)$, and
$$
\mathcal{C}_i^{(1)}(\theta)\cdot \mathcal{C}_j^{(1)}(\theta)^{\mathrm{T}}=\left\{
\begin{array}{ll}
0, & \mathrm{if}\,\ i=j,\\
1, & \mathrm{otherwise}.
\end{array}
\right.
$$
\end{lemma}
Proof.
Firstly, if $m\ge r+1$ we have $\mathcal{C}_j^{(r)}(\theta^{p^{m}})=\mathcal{C}_j^{(r)}(1)=(0,0,\ldots,0)$  since each $D_l^{(r)}(x)$ has $p-1$ many terms and $D_l^{(r)}(\theta^{p^{m}})=D_l^{(r)}(1)=p-1=0$ for all $0\le l<p^r$.
 Hence, for all $0\le i,j<p^r$ we have
$$\mathcal{C}_i^{(r)}(\theta)\cdot \mathcal{C}_j^{(r)}(\theta^{p^{m}})^{\mathrm{T}}=\mathcal{C}_i^{(r)}(\theta)\cdot \mathcal{C}_j^{(r)}(1)^{\mathrm{T}}=0. $$

Secondly, for $0\le i,j<p^r$ and $0\le m \le r$, we note that $D_l^{(r)}=g^{l} D_0^{(r)}$ for all $l\ge 0$ since we always suppose $Q_{r}(g)=1$, then we calculate
\begin{eqnarray*}
\mathcal{C}_i^{(r)}(\theta)\cdot \mathcal{C}_j^{(r)}(\theta^{p^{m}})^{\mathrm{T}} & = &
D_{i}^{(r)}(\theta)D_{j}^{(r)}(\theta^{p^{m}})+D_{i+1}^{(r)}(\theta)D_{j+1}^{(r)}(\theta^{p^{m}})+ \\
&& \ldots+D_{i+p^r-1}^{(r)}(\theta)D_{j+p^r-1}^{(r)}(\theta^{p^{m}})\\
          & = & \sum\limits_{k=0}^{p^r-1}~\sum\limits_{u\in D_0^{(r)}}\theta^{ug^{i+k}}~\sum\limits_{v\in D_0}\theta^{vg^{j+k}p^{m}}\\
          & = & \sum\limits_{k=0}^{p^r-1}~\sum\limits_{u\in D_0^{(r)}}\theta^{ug^{i+k}}~\sum\limits_{w\in D_0}\theta^{uwg^{j+k}p^{m}}~~ ~~(\mathrm{we~ use~ } v=uw)\\
          & = & \sum\limits_{k=0}^{p^r-1}~\sum\limits_{u\in D_0^{(r)}}~\sum\limits_{w\in D_0^{(r)}}\theta^{ug^{j+k}(g^{i-j}+wp^{m})}\\
          & = & \sum\limits_{w\in D_0^{(r)}}~\sum\limits_{z\in \Z_{p^{r+1}}^*}\gamma_w^z ~~(\mathrm{we~ use~ } z=ug^{j+k}, \gamma_w=\theta^{g^{i-j}+wp^{m}})\\
          & = & \sum\limits_{w\in D_0^{(r)}}~\sum\limits_{l=0}^{p^r-1}D_l^{(r)}(\gamma_w).
\end{eqnarray*}
Now we need to determine $\mathrm{ord}(\gamma_w)$, the order of $\gamma_w$ above for each $w\in D_0^{(r)}$. We note that $\mathrm{ord}(\gamma_w)|p^{r+1}$ since $\theta$ is a primitive
$p^{r+1}$-th root of unity.

If $1\le m\le r$, we find that $p\nmid (g^{i-j}+wp^{m})$ for all $w\in D_0^{(r)}$ and hence $\mathrm{ord}(\gamma_w)= p^{r+1}$. So we get
$$
\sum\limits_{w\in D_0^{(r)}}~\sum\limits_{l=0}^{p^r-1}D_l^{(r)}(\gamma_w)=0
$$
by Lemma \ref{lemma3}. We finish the proof of the first claim.

Now we consider the case $m=0$. For those $w\in D_0^{(r)}$ with $\mathrm{ord}(\gamma_w)\neq p$ we get
$$
\sum\limits_{l=0}^{p^r-1}D_l^{(r)}(\gamma_w)=0
$$
by Lemma \ref{lemma3} again. While in this case $(m=0)$, we show below that there exists $w\in D_0^{(r)}$ such that $\mathrm{ord}(\gamma_w)= p$ if and only if $p^{r-1}||(i-j)$.
That is, we need to find solutions $w\in D_0^{(r)}$ satisfying
$$
g^{i-j}+w \equiv l_0 p^{r} \pmod
{p^{r+1}}
$$
for some integer $l_0$ with
$1\le l_0<p$.
By (\ref{eq:add struct1}) and (\ref{eq:add
struct2}) we  get
\begin{equation}
\begin{split}\label{single-w}
0\equiv Q_{r}(w)&\equiv Q_{r}(-g^{i-j}+l_0p^{r})\\
& \equiv
Q_{r}(-g^{i-j})-l_0 p^{r-1} (-g^{i-j})^{-1}\\
&  \equiv
Q_{r}(-1)+(i-j)Q_{r}(g)-l_0 p^{r-1} (-g^{i-j})^{-1}\\
&  \equiv
(i-j)-l_0 p^{r-1} (-g^{i-j})^{-1}  \pmod {p^r}.
\end{split}
\end{equation}
Then for fixed $0\le i,j<p^r$, $l_0$ exists if and only if $p^{r-1}||(i-j)$. From (\ref{single-w}) we also
find that there is only one solution $l_0$ and hence only one $w$, written by $w_0$, such that $\mathrm{ord}(\gamma_{w_0})= p$,
in which case  we obtain by Lemma \ref{lemma3} again
$$
\sum\limits_{l=0}^{p^r-1}D_l^{(r)}(\gamma_{w_0})=1.
$$
So we conclude that
$$
\sum\limits_{w\in D_0^{(r)}}~\sum\limits_{l=0}^{p^r-1}D_l^{(r)}(\gamma_w)=\left\{
\begin{array}{ll}
1, & \mathrm{if}\,\ p^{r-1}||(i-j), \\
0, & \mathrm{otherwise},
\end{array}
\right.
$$
which finishes the proof of the second claim.

For the third claim, we can find the proof from \cite[Lemma 3]{C2014}.  ~\hfill $\square$\\

According to Lemma \ref{lemma02}, we remark that $u \bmod {p^{r+1}}\in D_{l~(\bmod {p^r})}^{(r)}$ if $u\in D_{l}^{(r+1)}$ for $r\ge 1$.
So together with Lemma \ref{lemma1}, we will use $\mathcal{C}_{l+p^r}^{(r)}(x)=\mathcal{C}_{l}^{(r)}(x)$ for any integer $l\ge 0$.

\begin{lemma}\label{single-def-pair}
Let $\mathfrak{r}\ge 2$ and $\beta \in \overline{\mathbb{F}}_{2}$ be a fixed primitive
$p^{\mathfrak{r}+1}$-th root of unity. Then for $0\le i<p^{\mathfrak{r}}$, the defining pair
of the binary sequence $(s^{(i)}_u)$ defined by
$$
s^{(i)}_u=\left\{
\begin{array}{ll}
1, & \mathrm{if}\,\ u\bmod {p^{\mathfrak{r}+1}} \in D_{i}^{(\mathfrak{r})},\\
0, & \mathrm{otherwise},
\end{array}
\right. \quad  u \ge 0,
$$
is $(G_{i}(x),\beta)$ with
$$
G_{i}(x)=\sum\limits_{k=1}^{p-1}x^{kp^{\mathfrak{r}}}+\sum\limits_{r=1}^{\mathfrak{r}}
\mathcal{C}_i^{(r)}(\beta^{p^{\mathfrak{r}-r}})\cdot \mathcal{C}_0^{(r)}(x^{p^{\mathfrak{r}-r}})^{\mathrm{T}}.
$$
\end{lemma}
Proof.
For $u=0$, we have
\begin{equation*}
\begin{split}
G_i(\beta^0)=G_i(1)&=\sum\limits_{k=1}^{p-1}1+\sum\limits_{r=1}^{\mathfrak{r}}
\mathcal{C}_i^{(r)}(\beta^{p^{\mathfrak{r}-r}})\cdot \mathcal{C}_0^{(r)}(1)^{\mathrm{T}}\\
&=(p-1)+0=0=s^{(i)}_0.
\end{split}
\end{equation*}

For $u= u'p^{m}$ with $\gcd(u', p)=1$ and $1\le m\le \mathfrak{r}$, we also suppose $u' \bmod p^{\mathfrak{r}+1}\in D_j^{(\mathfrak{r})}$ for some $j$,
then we derive by Lemmas \ref{lemma1} and \ref{inner-product}
\begin{eqnarray*}
G_{i}(\beta^u)& =&\sum\limits_{k=1}^{p-1}\beta^{kp^{\mathfrak{r}+m}u'}+\sum\limits_{r=1}^{\mathfrak{r}}
\mathcal{C}_i^{(r)}(\beta^{p^{\mathfrak{r}-r}})\cdot \mathcal{C}_0^{(r)}(\beta^{u'p^mp^{\mathfrak{r}-r}})^{\mathrm{T}}\\
             & = &p-1+\sum\limits_{r=1}^{\mathfrak{r}}
\mathcal{C}_i^{(r)}(\theta)\cdot \mathcal{C}_0^{(r)}(\theta^{u'p^m})^{\mathrm{T}}   ~~~~(\mathrm{we~ use~ } \theta=\beta^{p^{\mathfrak{r}-r}})\\
              & = & \sum\limits_{r=1}^{\mathfrak{r}}
\mathcal{C}_i^{(r)}(\theta)\cdot \mathcal{C}_j^{(r)}(\theta^{p^m})^{\mathrm{T}}=0=s^{(i )}_u.
\end{eqnarray*}

For $u\in D_{j}^{(\mathfrak{r})}$ with $0\le j<p^{\mathfrak{r}}$, we have by Lemma \ref{lemma1}
 \begin{eqnarray*}
G_{i}(\beta^u)& =&\sum\limits_{k=1}^{p-1}\beta^{kup^{\mathfrak{r}}}+\sum\limits_{r=1}^{\mathfrak{r}}
\mathcal{C}_i^{(r)}(\beta^{p^{\mathfrak{r}-r}})\cdot \mathcal{C}_0^{(r)}(\beta^{up^{\mathfrak{r}-r}})^{\mathrm{T}}\\
                  & = &1+\sum\limits_{r=1}^{\mathfrak{r}}
\mathcal{C}_i^{(r)}(\theta)\cdot \mathcal{C}_0^{(r)}(\theta^{u})^{\mathrm{T}}   ~~~~(\mathrm{we~ use~ } \theta=\beta^{p^{\mathfrak{r}-r}})\\
              & = &1+\sum\limits_{r=1}^{\mathfrak{r}}
\mathcal{C}_i^{(r)}(\theta)\cdot \mathcal{C}_j^{(r)}(\theta)^{\mathrm{T}}.
\end{eqnarray*}
We will proceed the proof by using the second and third claim in Lemma \ref{inner-product}.

If $p\nmid (i-j)$, then $i\not\equiv j \pmod {p^r}$ for all $r\ge 1$.  Hence we get
\begin{equation*}
\begin{split}
G_{i}(\beta^u)& =1+\mathcal{C}_i^{(1)}(\theta)\cdot \mathcal{C}_j^{(1)}(\theta)^{\mathrm{T}}+\sum\limits_{r=2}^{\mathfrak{r}}
\mathcal{C}_i^{(r)}(\theta)\cdot \mathcal{C}_j^{(r)}(\theta)^{\mathrm{T}}\\
&=1+1+0=0=s^{(i )}_u.
\end{split}
\end{equation*}
If $p^n||(i-j)$ for some $1\le n<\mathfrak{r}$, which indicates $i\equiv j \pmod {p^r}$ for all $1\le r\le n$ but $i\not\equiv j \pmod {p^r}$ for all $r> n$, then we get
\begin{equation*}
\begin{split}
G_{i}(\beta^u)& =1+\mathcal{C}_i^{(n+1)}(\theta)\cdot \mathcal{C}_j^{(n+1)}(\theta)^{\mathrm{T}}+\sum\limits_{\stackrel{r=1}{r\neq n+1}}^{\mathfrak{r}}
\mathcal{C}_i^{(r)}(\theta)\cdot \mathcal{C}_j^{(r)}(\theta)^{\mathrm{T}}\\
&=1+1+0=0=s^{(i )}_u.
\end{split}
\end{equation*}
Finally if $i=j$, we get
$$
G_{i}(\beta^u)=1+\sum\limits_{r=1}^{\mathfrak{r}}
\mathcal{C}_i^{(r)}(\theta)\cdot \mathcal{C}_j^{(r)}(\theta)^{\mathrm{T}}=1+0=s^{(i )}_u.
$$

Putting everything together, we get $s^{(i )}_u=G_{i }(\beta^u)$ for all $u\ge 0$ and complete
the proof. ~\hfill $\square$\\

Applying Lemma \ref{single-def-pair}, one can get the following important statement.
\begin{theorem}\label{def-poly}
 Let $\mathfrak{r}\ge 2$ and $\beta \in \overline{\mathbb{F}}_{2}$ be a fixed primitive
$p^{\mathfrak{r}+1}$-th root of unity. Then the defining polynomial $G(x)$ (corresponding to $\beta$)
of the binary sequence $(e_u)$ defined in (\ref{binarythreshold}) or (\ref{other-def})
is
$$
G(x)=\frac{p^{\mathfrak{r}}-1}{2}\sum\limits_{k=1}^{p-1}x^{kp^{\mathfrak{r}}}+\sum\limits_{r=1}^{\mathfrak{r}}~\sum\limits_{i=(p^r+1)/2}^{p^r-1}
\mathcal{C}_i^{(r)}(\beta^{p^{\mathfrak{r}-r}})\cdot \mathcal{C}_0^{(r)}(x^{p^{\mathfrak{r}-r}})^{\mathrm{T}}.
$$
\end{theorem}
Proof. By  Lemma \ref{single-def-pair} we see that the defining polynomial $G(x)$
of $(e_u)$ is
$$
G(x)=\frac{p^{\mathfrak{r}}-1}{2}\sum\limits_{k=1}^{p-1}x^{kp^{\mathfrak{r}}}+\sum\limits_{i=(p^\mathfrak{r}+1)/2}^{p^\mathfrak{r}-1}~\sum\limits_{r=1}^{\mathfrak{r}}
\mathcal{C}_i^{(r)}(\beta^{p^{\mathfrak{r}-r}})\cdot \mathcal{C}_0^{(r)}(x^{p^{\mathfrak{r}-r}})^{\mathrm{T}}.
$$
On the other hand, re-arranging the following summation, we get
$$
\sum\limits_{i=0}^{p^r-1}
\mathcal{C}_i^{(r)}(\beta^{p^{\mathfrak{r}-r}})\cdot \mathcal{C}_0^{(r)}(x^{p^{\mathfrak{r}-r}})^{\mathrm{T}}=\sum\limits_{i=0}^{p^r-1}
D_i^{(r)}(\beta^{p^{\mathfrak{r}-r}})\cdot \sum\limits_{l=0}^{p^r-1}D_l^{(r)}(x^{p^{\mathfrak{r}-r}})=0
$$
by Lemma \ref{lemma3} since $\mathrm{ord}(\beta^{p^{\mathfrak{r}-r}})=p^{r+1}$. Then using this fact we get
$$
\sum\limits_{i=(p^\mathfrak{r}+1)/2}^{p^\mathfrak{r}-1}
\mathcal{C}_i^{(r)}(\beta^{p^{\mathfrak{r}-r}})\cdot \mathcal{C}_0^{(r)}(x^{p^{\mathfrak{r}-r}})^{\mathrm{T}}=\sum\limits_{i=(p^r+1)/2}^{p^r-1}
\mathcal{C}_i^{(r)}(\beta^{p^{\mathfrak{r}-r}})\cdot \mathcal{C}_0^{(r)}(x^{p^{\mathfrak{r}-r}})^{\mathrm{T}},
$$
since the subscript $i$ of $\mathcal{C}_{i}^{(r)}$ is reduced modulo $p^r$. We finish the proof. ~\hfill $\square$\\

For example, let $p=5$ and $\mathfrak{r}=3$, we have
\begin{eqnarray*}
G(x)&=&62\sum\limits_{k=1}^{4}x^{kp^{3}}+\sum\limits_{i=3}^{4}
\mathcal{C}_i^{(1)}(\beta^{p^{2}})\cdot \mathcal{C}_0^{(1)}(x^{p^{2}})^{\mathrm{T}}\\
&&~~~~~ +\sum\limits_{i=13}^{24}
\mathcal{C}_i^{(2)}(\beta^{p})\cdot \mathcal{C}_0^{(2)}(x^{p})^{\mathrm{T}}+\sum\limits_{i=63}^{124}
\mathcal{C}_i^{(3)}(\beta)\cdot \mathcal{C}_0^{(3)}(x)^{\mathrm{T}}\\
&=&\sum\limits_{i=3}^{4}
\mathcal{C}_i^{(1)}(\beta^{p^{2}})\cdot \mathcal{C}_0^{(1)}(x^{p^{2}})^{\mathrm{T}}+\sum\limits_{i=13}^{24}
\mathcal{C}_i^{(2)}(\beta^{p})\cdot \mathcal{C}_0^{(2)}(x^{p})^{\mathrm{T}}\\
&&~~~~~+\sum\limits_{i=63}^{124}
\mathcal{C}_i^{(3)}(\beta)\cdot \mathcal{C}_0^{(3)}(x)^{\mathrm{T}}\in \mathbb{F}_2[x].
\end{eqnarray*}
Write $\theta_1=\beta^{p^{2}}$,~ $\theta_2=\beta^{p}$, ~ $\theta=\beta$. Then $\theta_1$ (resp. $\theta_2, \theta_3$) is a  primitive
$p^{2}$-th (resp. $p^{3}$-th, $p^{4}$-th) root of unity. Below we compute two examples.

If $u\in D_{17}^{(3)}$, then we have $u\in D_{2}^{(1)}$ and $u\in D_{17}^{(2)}$ by Lemma \ref{lemma02}, hence we see that
\begin{eqnarray*}
G(\beta^u)&=&\sum\limits_{i=3}^{4}
\mathcal{C}_i^{(1)}(\beta^{p^{2}})\cdot \mathcal{C}_0^{(1)}(\beta^{up^{2}})^{\mathrm{T}}+\sum\limits_{i=13}^{24}
\mathcal{C}_i^{(2)}(\beta^{p})\cdot \mathcal{C}_0^{(2)}(\beta^{up})^{\mathrm{T}}\\
&&~~~~~+\sum\limits_{i=63}^{124}
\mathcal{C}_i^{(3)}(\beta)\cdot \mathcal{C}_0^{(3)}(\beta^u)^{\mathrm{T}}\\
&=&\sum\limits_{i=3}^{4}
\mathcal{C}_i^{(1)}(\theta_1)\cdot \mathcal{C}_{2}^{(1)}(\theta_1)^{\mathrm{T}}+\sum\limits_{i=13}^{24}
\mathcal{C}_i^{(2)}(\theta_2)\cdot \mathcal{C}_{17}^{(2)}(\theta_2)^{\mathrm{T}}\\
&&~~~~~+\sum\limits_{i=63}^{124}
\mathcal{C}_i^{(3)}(\theta_3)\cdot \mathcal{C}_{17}^{(3)}(\theta_3)^{\mathrm{T}}\\
&=&(1+1)+1+ (1+1+1) \qquad\quad (\mathrm{by~ Lemma~ 4})\\
&=&0=e_u.
\end{eqnarray*}
If $u\in D_{85}^{(3)}$, then we have $u\in D_{0}^{(1)}$ and $u\in D_{10}^{(2)}$ by Lemma \ref{lemma02} again,   we get similarly
\begin{eqnarray*}
G(\beta^u)
&=&\sum\limits_{i=3}^{4}
\mathcal{C}_i^{(1)}(\theta_1)\cdot \mathcal{C}_{0}^{(1)}(\theta_1)^{\mathrm{T}}+\sum\limits_{i=13}^{24}
\mathcal{C}_i^{(2)}(\theta_2)\cdot \mathcal{C}_{10}^{(2)}(\theta_2)^{\mathrm{T}}\\
&&~~~~~+\sum\limits_{i=63}^{124}
\mathcal{C}_i^{(3)}(\theta_3)\cdot \mathcal{C}_{85}^{(3)}(\theta_3)^{\mathrm{T}}\\
&=&(1+1)+(1+1)+ 1=1=e_u.
\end{eqnarray*}

\section{Trace representation}

The trace representation plays an important role in sequence design. The trace function from $\F_{2^n}$ to $\F_{2^k}$ is
defined by
$$
\mathrm{Tr}^n_k(x)=x+x^{2^k}+x^{2^{2k}}+\ldots+x^{2^{(\frac{n}{k}-1)k}}.
$$
For $a,b\in\F_{2^k}$ and $x,y\in\F_{2^n}$, we have
$\mathrm{Tr}^n_k(ax+by)=a\mathrm{Tr}^n_k(x)+b\mathrm{Tr}^n_k(y)$. We
refer the reader to \cite{Lidl1983,GG} for details on the trace function.
The trace representations of many famous sequences, such as Legendre and Jacobi sequences and their generalizations, have
been studied in literature \cite{DGS,DGS09,DGSY}.

\begin{lemma}\label{lambda}
We suppose that $2^{p-1}\not\equiv 1 \pmod {p^2}$. If the order of 2 modulo $p$ is $\lambda$, then the order of 2 modulo $p^r$ is $\lambda p^{r-1}$ for $r\ge 2$.
\end{lemma}
Proof. Let $2^{\lambda}=1+k_0p$ for some integer $k_0$, since $\lambda$ is the order of 2 modulo $p$. We have
$$
2^{\lambda p^{r-1}}\equiv (1+k_0p)^{p^{r-1}}\equiv 1 \pmod {p^r}.
$$
According to the following two claims, we prove the desired result.

Claim 1. $2^{\lambda_1 p^{r-1}}\not\equiv  1 \pmod {p^r}$ for $\lambda_1<\lambda$ with  $\lambda_1|\lambda$.
\newline
(proof of Claim 1.) Since otherwise, we have $2^{\lambda_1} \equiv  1 \pmod {p}$, which contradicts to the condition that $\lambda$ is the order of 2 modulo $p$.

Claim 2.   $2^{\lambda p^{r-2}}\not\equiv  1 \pmod {p^r}.$
\newline
(proof of Claim 2.) We note first that $k_0\not \equiv 0 \pmod p$ since
$$
2^{p-1}\equiv (2^{\lambda})^{\frac{p-1}{\lambda}}\equiv (1+k_0p)^{\frac{p-1}{\lambda}}\equiv 1+\frac{p-1}{\lambda}k_0p \not\equiv 1 \pmod {p^2}.
$$
Then we have
$$
2^{\lambda p^{r-2}}\equiv (1+k_0p)^{p^{r-2}}\equiv 1 + k_0p^{r-1} \not \equiv 1\pmod {p^r}.
$$
We finish the proof. ~\hfill $\square$

\begin{lemma}\label{trace-single}
We suppose that $2^{p-1}\not\equiv 1 \pmod {p^2}$. Let $\lambda$ be the order of 2 modulo $p$ and $Q_r(g)=1$ for a (fixed) primitive root $g$ modulo $p^{r+1}$ for $r\ge 1$ as before.
We have
$$
D_l^{(r)}(x)=\sum\limits_{j=0}^{\frac{p-1}{\lambda}-1}\mathrm{Tr}^{\lambda p^r}_{p^r}\left(x^{g^{jp^r+l}}\right), ~~ l\ge 0.
$$
\end{lemma}
Proof. According to Lemma \ref{lambda}, we write
$$
U^{(r)}=\{2^{jp^r} \pmod {p^{r+1}} : 0\le j< \lambda\}\subseteq \mathbb{Z}_{p^{r+1}}^*.
$$
It is clear that $U^{(r)}$  is  a subgroup of $D_0^{(r)}$ due to $Q_r(2^{jp^r})\equiv jp^rQ_r(2) \equiv 0 \pmod {p^{r}} $ for $0\le j< \lambda$.
Then we divide  $D_0^{(r)}$ into $(p-1)/\lambda$ many subsets
$$
U^{(r)}, ~ g^{p^r}U^{(r)}, \ldots, g^{(\frac{p-1}{\lambda}-1)p^r}U^{(r)}.
$$
Now applying
$$
U^{(r)}(x)=\sum\limits_{u\in U^{(r)}} x^u=\mathrm{Tr}^{\lambda p^r}_{p^r}\left(x\right)\in \mathbb{F}_2[x],
$$
we derive
$$
D_0^{(r)}(x)=\sum\limits_{j=0}^{\frac{p-1}{\lambda}-1}\mathrm{Tr}^{\lambda p^r}_{p^r}\left(x^{g^{jp^r}}\right).
$$
Then the desired result follows from the fact that $D_l^{(r)}=g^lD_0^{(r)}$ for $l\ge 0$.
~\hfill $\square$

\begin{theorem}\label{trace-e}
 Let $\mathfrak{r}\ge 2$ and $\beta \in \overline{\mathbb{F}}_{2}$ be a fixed primitive
$p^{\mathfrak{r}+1}$-th root of unity.  Let $g$ be a (fixed) primitive root  modulo $p^{\mathfrak{r}+1}$ such that $Q_{\mathfrak{r}}(g)=1$. Let $\lambda$ be the order of $2$ modulo $p$. If $2^{p-1}\not\equiv 1 \pmod {p^2}$, then the trace representation of  $(e_u)$ defined in
(\ref{binarythreshold}) is
$$
e_u=\frac{p^{\mathfrak{r}}-1}{2}\sum\limits_{k=0}^{\frac{p-1}{\lambda}-1}\mathrm{Tr}^{\lambda}_1(\beta^{up^{\mathfrak{r}}g^k})+\sum\limits_{r=1}^{\mathfrak{r}}~\sum\limits_{l=0}^{p^r-1} \eta_l^{(r)}~\sum\limits_{j=0}^{\frac{p-1}{\lambda}-1}\mathrm{Tr}^{\lambda p^r}_{p^r}\left(\beta^{u p^{\mathfrak{r}-r}g^{jp^r+l}}\right),
$$
where
\begin{equation}\label{eta}
\eta_l^{(r)}=\sum\limits_{i=(p^r+1)/2}^{p^r-1}D_{i+l}^{(r)}(\beta^{p^{\mathfrak{r}-r}}).
\end{equation}
\end{theorem}
Proof. From Theorem \ref{def-poly}, we re-write the defining polynomial  $G(x)$  of $(e_u)$ as
\begin{eqnarray}\label{Gx}
G(x) = \frac{p^{\mathfrak{r}}-1}{2}\sum\limits_{k=1}^{p-1}x^{kp^{\mathfrak{r}}}+\sum\limits_{r=1}^{\mathfrak{r}}~\sum\limits_{l=0}^{p^r-1}\eta_l^{(r)}D_l^{(r)}(x^{p^{\mathfrak{r}-r}}),
\end{eqnarray}
where $\eta_l^{(r)}$ is defined in (\ref{eta}). The trace representation of $D_l^{(r)}(x^{p^{\mathfrak{r}-r}})$ is given in Lemma \ref{trace-single}.
We remark that $g$ is also a primitive root  modulo $p^{r}$ and $Q_{r}(g)=1$ for all $1\le r\le \mathfrak{r}+1$ since we suppose that $g$ is a  primitive root  modulo $p^{\mathfrak{r}+1}$ such that $Q_{\mathfrak{r}}(g)=1$.
So we only need to describe  $\sum\limits_{k=1}^{p-1}x^{kp^{\mathfrak{r}}}$ by using trace function.

Since $\lambda$ is the order of $2$ modulo $p$ and $g$ is also a primitive root modulo $p$, we have
$$
\Z_p^*=\bigcup\limits_{k=0}^{\frac{p-1}{\lambda}-1}g^k \langle 2\rangle,
$$
where $\langle 2\rangle=\{1,2,2^2,\ldots,2^{\lambda-1}\}$ generated by $2$ modulo $p$ is a subgroup of $\Z_p^*$. Hence we derive
$$
\sum\limits_{k=1}^{p-1}x^{kp^{\mathfrak{r}}}=\sum\limits_{k=0}^{\frac{p-1}{\lambda}-1}\mathrm{Tr}^{\lambda}_1(x^{p^{\mathfrak{r}}g^k}).
$$
We complete the proof. ~\hfill $\square$\\

For the case of $2^{p-1}\equiv 1 \pmod {p^2}$, we see that  the order of 2 modulo $p^r$ is not always $\lambda p^{r-1}$, where  $\lambda$ is the order of 2 modulo $p$.
For example, for $p=1093$, the experimental result shows that the order of 2 modulo $p^r$ is $\lambda=364$ for $r=1$ or $2$ and the order of 2 modulo $p^r$ is $\lambda p^{r-2}$ for $r\ge 3$.

In fact, for any such $p$ (i.e., satisfying $2^{p-1}\equiv 1 \pmod {p^2}$), if $\lambda$ is the order of 2 modulo $p^r$ for all $1\le r\le t_0$ with a maximal integer $t_0$, then the order of 2 modulo $p^r$ is $\lambda p^{r-t_0}$ for all $r\ge t_0+1$ by using a similar proof of Lemma \ref{lambda}. In terms of
$$
U^{(r)}=\{2^{j} \pmod {p^{r+1}} : 0\le j< \lambda\}\subseteq D_0^{(r)}, ~~~ r<  t_0
$$
and
$$
U^{(r)}=\{2^{jp^{r+1-t_0}} \pmod {p^{r+1}} : 0\le j< \lambda\}\subseteq D_0^{(r)}, ~~~ r\ge t_0,
$$
one can apply the idea of Lemma \ref{trace-single} to describing the trace of each $D_l^{(r)}(x)$ and hence the defining polynomial $G(x)$  of $(e_u)$ without any difficulties.

We finally remark that such primes $p$, which are called Wieferich primes,  are very rare. To date the only known such primes are
$p=1093$ and $p=3511$ and it was  reported that there are no new
such primes $p< 4\times 10^{12}$, see \cite{CDP1997}.

\section{Final remarks}

In this manuscript, we give the trace representation of a family of binary threshold sequences derived from Euler quotients by determining
the corresponding defining polynomials.

The defining polynomial of a sequence plays an important role in cryptography. It is closely related to the  linear complexity of the sequence.
We recall that the
\emph{linear complexity} $L((e_u))$  is the least order $L$ of a linear
recurrence relation over $\mathbb{F}_2$
$$
e_{u+L} = c_{L-1}e_{u+L-1} +\ldots +c_1e_{u+1}+ c_0e_u\quad
\mathrm{for}\,\ u \geq 0,
$$
which is satisfied by $(e_u)$ and where $c_0=1, c_1, \ldots,
c_{L-1}\in \mathbb{F}_2$. For a sequence to be cryptographically strong, its linear complexity
should be large and at least a half of the period  according to  the Berlekamp-Massey
algorithm \cite{Massey}. From \cite{B} or \cite[Theorem 6.3]{GG}, the linear complexity of $(e_u)$ equals the number of nonzero coefficients of the defining polynomial $G(x)$, i.e., the \emph{Hamming weight} of $G(x)$.

According to the proof of \cite[Lemma 6]{DCH-IPL}, we see that $\eta_l^{(r)}\neq 0$ in (\ref{eta}) for all $1\le r\le \mathfrak{r}$ and $0\le l<p^r$. Hence if $2^{p-1} \not\equiv 1 \pmod {p^2}$, we get
by computing the Hamming weight of $G(x)$ in (\ref{Gx})
\begin{eqnarray*}
L((e_u))& = &\sum\limits_{r=1}^{\mathfrak{r}}p^r(p-1)+(p-1)\epsilon\left(\frac{p^{\mathfrak{r}}-1}{2}\right)\\
&=& p^{\mathfrak{r}+1}-p+(p-1)\epsilon\left(\frac{p^{\mathfrak{r}}-1}{2}\right)\\
&=&\left\{
\begin{array}{ll}
p^{\mathfrak{r}+1}-p, & \mathrm{if}\,\ p \equiv 1 \pmod 4, \\
p^{\mathfrak{r}+1}-p, & \mathrm{if}\,\ p \equiv 3 \pmod 4 ~\mathrm{and}~ \mathfrak{r}~ \mathrm{is~ even},\\
p^{\mathfrak{r}+1}-1, & \mathrm{if}\,\  p \equiv 3 \pmod 4 ~\mathrm{and } ~\mathfrak{r}~ \mathrm{is~ odd},\\
\end{array}
\right.
\end{eqnarray*}
which has been proved in \cite[Theorem 1]{DCH-IPL}. The notation $\epsilon\left(\frac{p^{\mathfrak{r}}-1}{2}\right)$ above satisfies
\begin{eqnarray*}
\epsilon\left(\frac{p^{\mathfrak{r}}-1}{2}\right)& = &\left\{
\begin{array}{ll}
0, & \mathrm{if}\,\ \frac{p^{\mathfrak{r}}-1}{2} ~ \mathrm{is~ even},\\
1, & \mathrm{otherwise}.
\end{array}
\right.
\end{eqnarray*}

 %Selecting different $\mathcal{I}\subseteq \mathbb{Z}_{p^\mathfrak{r}}$, one can define different families of binary sequences.
%In particular, if one uses Jacobi symbol $\left(\frac{\cdot}{p^\mathfrak{r}}\right)$ to define
%$$
%\mathcal{I}=\left\{m :  m\in \mathbb{Z}_{p^\mathfrak{r}},  ~ \left(\frac{m}{p^\mathfrak{r}}\right)=1   \right\}
%$$

\section*{Acknowledgements}

Z. Chen was partially supported by the National Natural Science
Foundation of China under grant No.61373140. ~~ X. Du was partially supported by the National Natural Science
Foundation of China under grant 61202395 and the Program for New Century Excellent Talents in University (NCET-12-0620).


\begin{thebibliography}{99}

\bibitem{ADS}  T. Agoh, K.Dilcher, L. Skula, Fermat quotients for composite
moduli, J. Number Theory 66(1) (1997) 29-50.

\bibitem{AW} H. Aly  and A. Winterhof. Boolean functions derived from Fermat
quotients. Cryptogr. Commun. 3 (2011) 165--174.


\bibitem{B} R. E. Blahut. Transform techniques for error control codes.
IBM J. Res. Develop. 23 (1979)  299--315.



\bibitem{BFKS}
J. Bourgain, K. Ford, S. Konyagin and I. E. Shparlinski. On the
divisibility of Fermat quotients. Michigan Math. J. 59 (2010) 313--328.



\bibitem{C}
M. C. Chang. Short character sums with Fermat quotients. Acta Arith.
152 (2012) 23--38.


\bibitem{C2014} Z. X. Chen. Trace representation and linear complexity of binary
sequences derived from Fermat quotients. Sci. China Inf. Sci. (2014)  (to appear)


\bibitem{CD}  Z. X. Chen  and  X. N. Du. On the linear complexity of binary threshold sequences derived from
Fermat quotients. Des. Codes Cryptogr. 67 (2013) 317--323.



\bibitem{CG}Z. X. Chen  and  D. G\'{o}mez-P\'{e}rez. Linear complexity of
binary sequences derived from polynomial quotients. Sequences and Their Applications-SETA 2012, 181--189, Lecture Notes in Comput. Sci., 7280, Springer, Berlin, 2012.


\bibitem{CNW} Z. X. Chen, Z. H. Niu and C. H. Wu. On the $k$-error linear complexity of binary
sequences derived from polynomial quotients. http://arxiv.org/abs/1307.6626, 2013.



\bibitem{CHD2011}  Z. X. Chen, L. Hu, X. N. Du, Linear complexity of some binary sequences
derived from Fermat quotients, China Commun. 9(2) (2012) 105-108.


\bibitem{COW} Z. X. Chen, A. Ostafe, A. Winterhof, Structure of pseudorandom numbers
derived from Fermat quotients, in: Proc. of WAIFI 2010, Lecture
Notes in Comput. Sci., vol. 6087, Springer-Verlag, Heidelberg, 2010,
pp.73-85.

\bibitem{CW11}  Z. X. Chen, A. Winterhof, On the distribution of pseudorandom numbers and vectors derived from
Euler-Fermat quotients, International Journal of Number Theory 8
(3)(2012) 631-641.




\bibitem{CW2}Z. X. Chen  and A.  Winterhof. Additive character sums of
polynomial quotients. Theory and Applications of Finite Fields-Fq10,
67--73, Contemp. Math., 579, Amer. Math. Soc., Providence, RI, 2012.



\bibitem{CW3} Z. X. Chen and A. Winterhof. Interpolation of Fermat quotients. SIAM J. Discr. Math. 28 (2014) 1--7.



\bibitem{CDP1997}  R. Crandall, K. Dilcher and C. Pomerance. A search for Wieferich
and Wilson primes. Math. Comp. 66 (217) (1997) 433--449.




\bibitem{DGS}
Z. D. Dai, G. Gong and  H. Y. Song. Trace representation and linear complexity of binary $e$-th residue sequences. Int'l Workshop on Coding and Cryptography-WCC 2003, 121--133, Versailles, France, 2003.



\bibitem{DGS09}
Z. D. Dai, G. Gong and  H. Y. Song. A trace representation of binary Jacobi sequences. Discrete Math. 309 (2009)  1517--1527.


\bibitem{DGSY}Z. D. Dai, G. Gong, H. Y. Song and  D. F. Ye. Trace representation and linear complexity of binary $e$-th power residue sequences of period $p$. IEEE Trans. Inform. Theory 57 (2011)  1530--1547.




\bibitem{DCH-IPL} X. N. Du, Z. X. Chen,  L. Hu, Linear complexity of binary sequences derived from
Euler quotients with prime-power modulus, Information Processing
Letters, 112(12) (2012) 604-609.


\bibitem{DKC}  X. N. Du,  A. Klapper, Z. X. Chen, Linear complexity of pseudorandom sequences generated by
Fermat quotients and their generalizations, Information Processing
Letters 112(6) (2012) 233-237.


\bibitem{ErnMet2} R. Ernvall,  T. Mets{\"a}nkyl{\"a},
On the $p$-divisibility of Fermat quotients,  Math. Comp. 66(219)
(1997) 1353-1365.




\bibitem{GG} Golomb S W, Gong G. Signal Design for Good Correlation. Cambridge: Cambridge University Press, 2005




\bibitem{GW} D. G\'{o}mez-P\'{e}rez and A. Winterhof. Multiplicative character sums of
Fermat quotients and pseudorandom sequences. Period. Math. Hungar.
64 (2012) 161--168.


\bibitem{J} D. Jungnickel. Finite Fields: Structure and Arithmetics. Bibliographisches Institut, Mannheim, 1993.



\bibitem{Lidl1983} R. Lidl, H. Niederreiter, Finite Fields, Addison-Wesley, Reading, MA,
1983.


\bibitem{Massey} J. L. Massey,  Shift register synthesis and BCH decoding, IEEE Trans. Inform.
Theory 15(1) (1969) 122-127.

\bibitem{Nathanson} M. B. Nathanson. Elementary Methods in Number Theory. Graduate
Texts in Mathematics, 195. Springer-Verlag, New York, 2000.

\bibitem{OS}  A. Ostafe, I.E. Shparlinski, Pseudorandomness and dynamics of Fermat
quotients, SIAM J. Discr. Math. 25(1) (2011) 50-71.


\bibitem{Sha}M. Sha, The arithmetic of Carmichael quotients, http://arxiv.org/abs/1108.2579v5, 2013.


\bibitem{S}I. E. Shparlinski, Character sums with Fermat quotients, Quart. J. Math. 62(4) (2011) 1031-1043.



\bibitem{S2010} I. E. Shparlinski,  Bounds of multiplicative character sums with
Fermat quotients of primes, Bull. Aust. Math. Soc. 83(3) (2011)
456-462 .


\bibitem{S2011} I. E. Shparlinski,  On the value set of Fermat
quotients, Proc. Amer. Math. Soc. 140(4) (2012) 1199-1206.


\bibitem{S2011b} I. E. Shparlinski,  Fermat quotients: Exponential sums, value set and primitive
roots, Bull. Lond. Math. Soc.  43(6) (2011) 1228-1238.

\bibitem{SW}
I. E. Shparlinski  and A. Winterhof. Distribution of values of
polynomial Fermat quotients. Finite Fields Appl. 19 (2013)
93--104.


\bibitem{W} A. Winterhof, Linear complexity and related complexity measures, In
selected topics in information and coding theory, World
 Scientific, 2010, pp.3-40.





\end{thebibliography}
\end{document}